# FEATURE SELECTION APPROACHES FOR OPTIMISING MUSIC EMOTION RECOGNITION METHODS

Le Cai[1], Sam Ferguson[2], Haiyan Lu[2] and Gengfa Fang[1]

[1]School of Electrical and Data Engineering
[2]School of Computer Science, Faculty of Engineering & IT, University of Technology Sydney, Sydney, NSW, 2007, Australia.

## ABSTRACT

*The high feature dimensionality is a challenge in music emotion recognition. There is no common consensus on a relation between audio features and emotion. The MER system uses all available features to recognize emotion; however, this is not an optimal solution since it contains irrelevant data acting as noise. In this paper, we introduce a feature selection approach to eliminate redundant features for MER. We created a Selected Feature Set (SFS) based on the feature selection algorithm (FSA) and benchmarked it by training with two models, Support Vector Regression (SVR) and Random Forest (RF) and comparing them against with using the Complete Feature Set (CFS). The result indicates that the performance of MER has improved for both Random Forest (RF) and Support Vector Regression (SVR) models by using SFS. We found using FSA can improve performance in all scenarios, and it has potential benefits for model efficiency and stability for MER task.*

## 1. INTRODUCTION

Music has become an indispensable part of people's lives. It plays a vital role in our world. We use music in almost every field, such as public places, entertainment, and even therapy. As the technology grows, the widespread adoption of digital audio formats, especially MP3, music distribution has become very efficient and seamless. The primary method of music consumption has shifted from retail stores to online and internet-based distribution channels. Subscription services had now become popular where the consumers now have access to much larger libraries than when albums were purchased individually. Traditional approaches to managing digital music libraries using of embedded metadata are no longer sufficient to deal with such a large database since the text cannot fully convey the expression of the musical content [1] [2], therefore the content-based music retrieval system can be ideal to handle this task more efficiency and opens a new perspective to discover music.

The core component of music is the expression of emotion, it can evoke powerful emotional responses in listeners [3] [4]. Determining the emotional content of musical signals relies not only on signal processing and machine learning but also on other study areas, such as psychology, auditory perception, and music theory [5] [6]. However, the emotion label retrieved by machine learning is still accurate enough to be beneficial for music content organization tasks and can also benefit other fields such as recommendation systems, music content retrieval, automatic music composing, and psychotherapy, and therefore it has become an area that has expanded significantly in the past decade. Many researchers are developing a new way to automate recognizing emotion labels, and thus Music Emotion Recognition has made significant progress and has become a research hotspot in the academic field and industrial application [7] [8] [9].

MER is a technique of extracting the emotion labels from an audio signal. Existing MER approaches require using many features to train a recognition model. This may employ excessive computational powers not only do the large number of features need to be computed for each audio signal, the training of the model is also made more complex by using the large number of input features. Due to this fact, MER is generally considered as a challenging task and requires high hardware applicability on device selection. However, not all audio features contain information equally relevant to emotion recognition. Firstly, Mion and De Poli suggest that some features may be redundant for music emotion recognition due to noise created by irrelevant data [10]. Second, there is no common consensus on which audio features are useful for MER, and therefore most approaches rely on dimension reduction techniques to construct a large potential feature set and expect it can be useful for MER [11] [12]. Third, there are few studies have focused on feature selection algorithms (FSAs) based approach for MER, Instead, a lot of research focuses on more advanced machine learning techniques, such as deep learning or reinforcement learning to automate the feature weighting in a black box type of approach. Therefore, to find out what features are useful is crucial, using the relevant features may improves the performance to the overall MER task.

The core motivation of this study is to find out how we can use feature selection to improve MER performance without knowing in advance which features are useful for music emotion recognition, thus we created a subset by excluding the redundant features from the complete data set. The paper is organized in the following structure.

- In Section I, we evaluate the existing studies of general musical features, MER approaches and the challenges particularly related to MER model.
- In Section II, we introduce the feature selection algorithms that can be useful for selecting MER features and the machine learning models that can be used for the recognition task.
- In Section III, we go through the experiment steps of how we created the subset, and benchmark it with the baseline model of using the full DEAM set.
- In Section IV, we discuss the benchmark results and insights of how FSAs can possibly benefit the MER and suggest future work directions.

## 2. RELATED WORK

### 2.1. Music Emotion Recognition

Music Emotion Recognition (MER) is a music information retrieval technique that extracts the emotion label from the raw audio features. MER extracts the emotion from various features, such as chroma, MFCC, spectral, etc. These features are analysed and processed through a computational model to get the emotion label.

The earliest study was done by Hevner in 1935. She studied the relationship between music and emotion by surveying people's feedback on the emotion of what they hear. Throughout her experiment, she collected 66 adjectives from the participants. These terms can be categorised into eight groups, providing the initial insights for the MER emotion conceptual model. These descriptors soon become Hevner's adjective cycle later, which is the baseline for currently most MER emotional models [13].

The current MER models can mainly summarise in two approaches: Categorical and Dimensional [14]. The Categorical approach treats emotions as discrete values and clusters them into groupings of emotions under broad categories such as happy, sad, angry, and relaxed categories. There are few emotions clusters model available for the categorical approach. The most famous model used in the categorical approach is the Hevner's adjective cycle [1], which laid out the emotion clusters in a circle from negative to positive level in a cumulative way until all the clusters are placed shows in Figure 1.

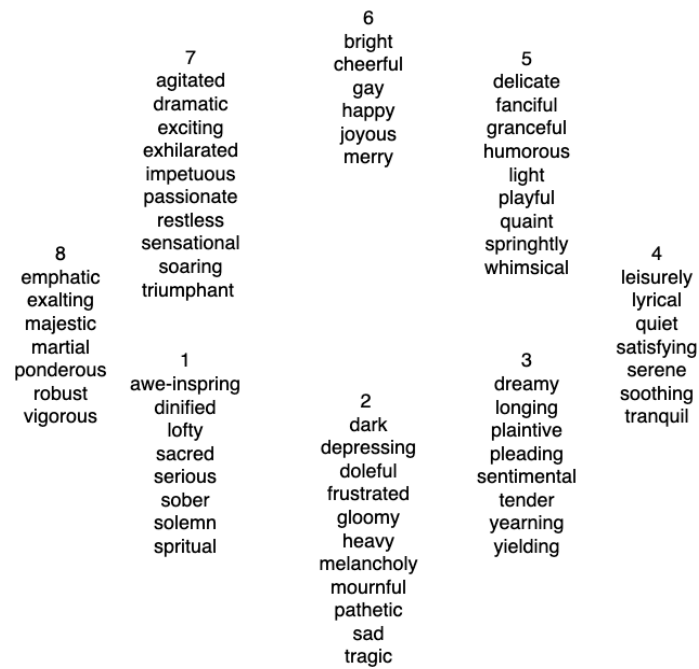

Figure 1. Hevner's adjective cycle

Hevner's model also has the advantage in regardless to listener's knowledge. Her study shows these labels are intuitive and consistent regardless of the listener's music training level [15]. The annotation can be collected from general public participants without acquiring expert knowledge, hence many MER studies still considered Hevner's model as the baseline approach preceding to emotion conceptions model [16] [17]. Another well-known model used widely in psychology is called The Six Universal Emotions developed by Ekman in 1982 [18]. It has a smaller amount of emotion clusters than Hevner's model. Ekman believes the universal emotion of humans can be classified into six unique clusters: Anger, Disgust, Fear, Happiness, Sadness, and Surprise. This model has the advantage across cultures and can be experienced universally by all human nature. The Six Universal Emotions has been used widely in the facial recognition task and obtained a great result in classifying the distinct emotion between different facial expressions. However, another study suggests this model may not be suitable for the MER task since it lacks some musical mood taxonomy (e.g., clam, soothing) [19]. Thus, Hevner's study is still considered the best-known methodical to conceptualise emotion in the MER field.

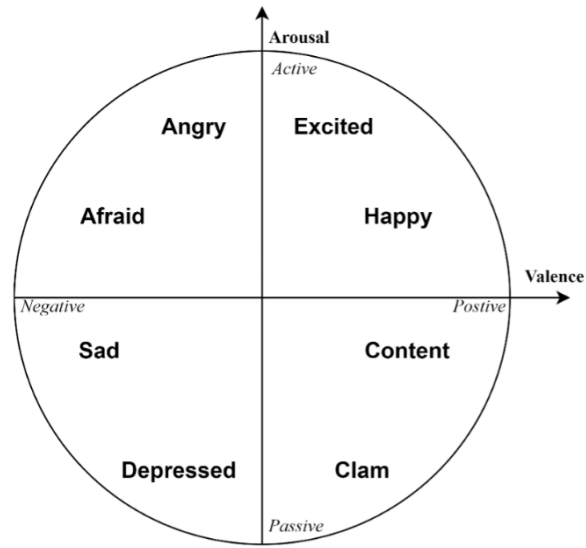

Figure 2. Valence and Arousal (VA) Model

The dimensional approach treats the emotion as continuous multidimensional space. The most noted model is Valence and Arousal (VA), created by Russell in 1980 who laid the foundation work for this approach [20]. The VA model creates a 2- dimensional space in a unit-circle, breaking down the emotions into two components: valence and arousal laid down it as a point in a plane as shown in Figure 2.

The arousal indicates the intensity ranging from intense to calm. Valence indicates the appraisal polarity range from negative to positive emotion. In this case, the emotion can be classified by each quadrant, setting with the appropriate boundary.

In contrast to the categorical based model, the dimension approach has serval advantages, firstly one common belief is that the class of categorical models are too small to convey the richness of emotion perceived by humans [21]. Conversely, continuous progressive refinement of the class can overwhelm the subjects. This has been considered impractical in psychological studies [22]. Second, using descriptive terms for emotion classes may have equivocalness because the languages may have an ambiguous understanding between different people [23]. This can cause bias in ground truth data collection and may inflect the training process of the MER model.

Moreover, the dimensional model is graphical intuitive and has the advantage of describing the correlation between emotional status and music features. It represents an emotional state in 2 components: Valence (V) and Arousal (A), corresponding to pitch (high or low) and tempo (fast or slow) in a coordinate to compare with other emotions. Furthermore, the dimensional model has excellent compatibility with adapting categorical models. Any label-type data can easily convert to dimensional models by eliminating regions on the quadrants. Therefore, this approach is generally considered the most common approach and it's used widely in MER task [21].

## 2.2. Musical Features

The way of people percept music is multidimensional. However, different emotional perceptions can associate with the various clues of acoustic features [24] [25]. For instance, arousal can usually relate to tempo, represented in slow to fast. Other features may also associate with arousal, such as loudness (loud/quiet), pitch (low/high) and timbre (bright/soft). Meanwhile, valance does not have a clear relationship like arousal does, but mode (major/ minor) and harmony (constant/dissonant) generally consider the valence's relevant features.

Schmidt et al. has conducted the work to find out the relation and performance of each feature to the music emotion [26]. Their work investigates the use of multiple acoustic features against to using them in combination. They found that the MFCCs and spectral contrast performs highest in individual features. However, the best overall result they obtain is by using the features in combination, showing that there is no dominant feature related to emotion. Their work suggests the emotion perception should not be dependent on using a single feature but on a combination of them. Table 1 describes a set of musical feature categories.

Table 1. Musical features category

| Category | Features |
|---|---|
| Energy | Loudness, RMS energy |
| Rhythm | Regularity, clarity, onset frequency, tempo |
| Melody | Pitch, chroma, tonality. |
| Timbre | MFCC, spectral shape, spectral contrast |

In the following section, we go through common audio acoustic features in their category and discuss how they relate to the music emotion.

- **Energy** – The energy is one of high co-related arousal feature, such as RMS energy and loudness. RMS is the root-mean-square measures the signal amplitude averaged over a time period. It's a metering tool measures the average loudness of an audio in a selected frame. Loudness measures the intensity of any given signal to the threshold of hearing. It's usually represented in dB ranging from 0 to infinity, however the threshold is generally considered below 140 dB is safe, anything about it can cause damage in hearing.
- **Rhythm** – Rhythm is the pattern of notes in the movement of sound, generally it mostly relates to the valance. E.g., a flowing or fluent rhythm usually associates with the positive valance, but in other hand a firm rhythm has the negative valance. There are several properties made up rhythm in music: rhythm strength, regularity, clarity, average onset frequency and tempo. Rhythm strength is the average intensity on a onset detection curve. It can compute by klapuri algorithm [27]. Rhythm regularity and clarity is the correlation between the peaks in onset detection curve, it measures how obvious of a rhythm in a music segment. Tempo indicates the number of beats per second through the music segment, it's calculated by the periodicity estimation from the onset detection curve.
- **Melody** – Melody is the key dimension of how human can recognize a music in our mind. This includes pitch, chroma and tonality features. Pitch identifies the frequency related in a scale, normally co-responding to the keys on a musical scale. Chroma or chromogram is a pitch class profile organize the entire spectrum into the distinct tones. It's a powerful visual representation tools for analysing the music melody. Springer Nature 2021 LATEX template 5 chromogram projects each pitch class into 12 bins, represent as 12 distinct semitones regardless of its absolute frequency. It can give us useful information for analysing the pitch distributions because the melody doesn't have much difference when playing in a different octave.

- **Timbre** – Timbre represents the sound quality, it provides information allows us to distinguish between different sound source, e.g., trumpet and a saxophone. this is extremely useful on determine what instruments are playing on a particular note because in general multiple instruments can play on a same note simultaneously to enhance a music piece.

### 2.3. Model complexity and feature selection

MER models require training before they can use to recognize music emotion. One assumes to improve MER performance is by using the most available music features for model training. However, this has been disproved by researchers, arguing that it can lead to creating more noise and decrease the recognition results [28] [29].

Zhang et al. suggest that using all existing features may not be efficient for MER tasks due to it creates redundant data lacks informative descriptors related to emotion [30]. Panda et al. argue that vibrato, tremolo, and articulation are expressive features related to music emotion [31] [32]. They extracted emotion expressive features and trained their model against to using the whole features, their results show that using the appreciate chosen features can boost the model performance, indicates that using selective features can be a possible approach to improve the MER performance.

Selecting the right audio features is the key to tackling this problem. Therefore, dimension reduction techniques need to be applied to the feature set. There are two primary methods for reducing the features, feature projection (FP) and feature selection (FS). Feature projection transforms the features from high dimension to low dimension by applying linear or non-linear approaches. Such as principal component analysis (PCA) and non-linear isometric feature mapping (ISOMAP). However, one issue with this approach is that the original features are transformed and modified. We cannot easily determine the importance of each original feature from the transformed version. On the other hand, the feature selection approach applies statistical analysis to the feature space, running the test with a combination of features to evaluate its weights and performance. Filters methods and wrapper methods are the two commonly used methodologies in FS. Filter methods, such as Relief, select the feature by analysing its statistics correlation based on training data, wrapper methods select feature relies on a learning algorithm, such as forward selection, backward elimination and recursive feature elimination, the evaluation of the feature is based on the selected model. Either of these methods are well developed and can effectively reduce the feature dimension.

## 3. METHODS

### 3.1. The VA Model

Before we dive into the feature selection work, we need a proper way to describe emotion. Currently, there are two models are mainly used in the MER task, dimensional and categorical. Dimensional models treat emotions as points within a continuous numerical axis. Russell and Thayer laid out the foundations for the dimensions of emotion. The most noted model used in the dimensional approach is Russell's and Thayer's model (also known as the VA Model), which organizes the emotional descriptors in four quadrants [33] [34] [35]. The model uses continuous value of valence and arousal to represent an emotion on a plane, as shown in Figure 2. The arousal indicates the intensity ranging from intense to calm. Valence indicates the appraisal polarity range from negative to positive emotion. The combination of these two values can construct a specific point in the coordinate. Multiple studies have confirmed that this model's validity performs a feasible practice and can efficiently describe the emotion.

The VA model can be represented in as a two-dimensional space, the emotion space coordinates can be reduced to an emotion category by simple Euclidean thresholding. In this case each emotion can spited into 2 components, valence (V) and arousal (A) as a coordinate of a point, valence represent the positivity measurement, range from -1 to 1, arousal indicates the intensity measurement of the emotion in same range of valence.

By introducing of the VA model, the MER can produce the emotion labels mapping from its audio feature input based on the coordinate of emotions in 4 quadrants. Showing in Figure 2.

The VA model has a distinct emotion categories, they can class by 4 quadrants value in [-1,1], [1,1], [-1,-1] and [-1,1]. In this case, we easily adapt the VA model to any categorical model by placing its emotion adjectives in the correct place to compatible with various of different categorical model. They can be achieved by eliminating the region of each quadrant, e.g., excited in $[0 < V < b1, b2 < A < 1]$, where b1 and b2 is the boundary of a emotion term, they can be adjusted by the experts according to the needs. In our case, we adapt Hevner's eight clusters of affective terms into our emotion plane [36] as we mentioned in related work, it's a well-known categorical approach in MER. Heaven's cluster has eight distinct emotion classes that vary in a cumulative way to contrast its neighbouring in opposites position. We can lay out these adjectives based on their clusters and neighbouring in a circle, shown in Figure 2. Each cluster is divided by half of the region in a quadrant. It makes up 2 clusters in a quadrant, fits up to 8 eights in total (4∗2 clusters per quadrant), and converts Hevner's clusters to our two-dimensional plane for the MER to use.

## 3.2. Data Acquisition

We use MediaEval DEAM dataset 1 in our experiment, the dataset is consisting of 1802 fully annotated songs with valence (V) and arousal (A) values and contains 260 audio features for each of them [37]. They are extracted by the openSMILE toolbox with a 500ms window size at a sampling rate of 44100Hz [38]. The label for each song is annotated by expert listeners, providing a 45s duration time sequence of dynamic information for valence and arousal. It provides standard deviation and means alongside the feature structures containing the song ID, music features and emotion labels. The distribution of dataset is shown in Figure 3.

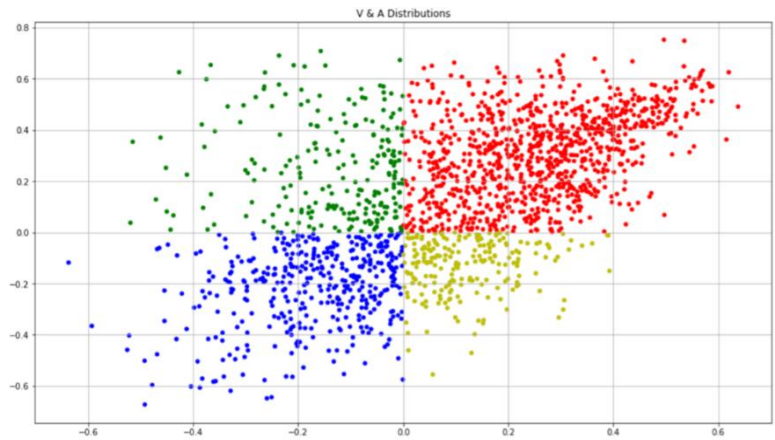

Figure 3. The data distribution

The data is categorized in 4 clusters, each colour represents the different emotion co-responding to its V, A values. Such as RED: happy, excited; GREEN: angry, afraid; BLUE: sad, depressed; YELLOW: clam, content. We adapt them into eight emotions clusters based on the model using our selected feature set and validation set separately to evaluate the result and benchmark the model performance.

### 3.3. Experimental Process

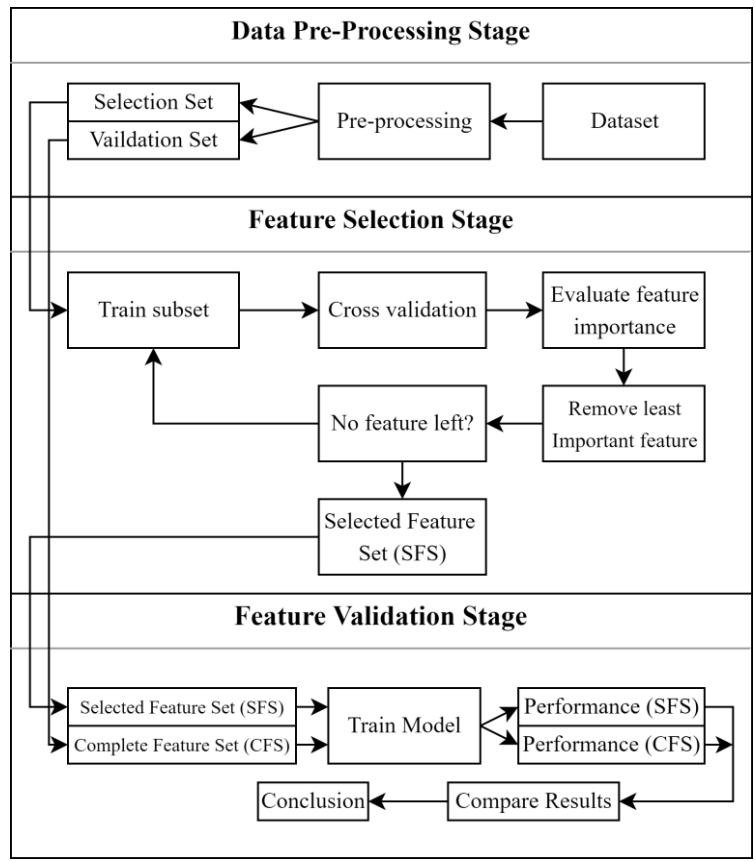

Figure 4. The Experiment Process

The overall process can be summaries into three main stages: data pre-processing, feature selection, and feature validation. Firstly, the data is pre-processed into two subsets, the selection set and the validation set, using the ratio of 70% to 30%.

Next, the selection set is used as an input to the feature selection algorithm which uses Recursive Feature Elimination (RFE), to determine an optimal number of the best features for undertaking model training. RFE recursively selects the high-ranking features, and then construct a selected feature set.

In the Validation & Benchmark stage, we use two different datasets to train the model, the selected feature set (generated from the previous RFE algorithm) and the validation set, for comparing & benchmarking the model's performance. We use two popular machine learning algorithms to benchmark our selected feature set, the Support Vector Regression (SVR) and Random Forest (RF). We use these two algorithms to train the model using our selected feature set and validation set separately to evaluate the result and benchmark the model performance.

### 3.4. Data Pre-Processing

The purpose of data pre-processing is to convert the data from the DEAM database to a standardized format since the benchmark regressor only supports vector-based inputs.

To optimize the variations between features, first the z-score normalization is applied for all the features from the original set. Next all-time sequence-based data are processed, ripping from 15s to 44.5s due to the limitation of the data sample length. After that we average all the values by the feature types, including its annotation of VA values from the label set and convert them into a suitable format for machine learning model to use.

The experiment mainly consists with two datasets, The Complete Feature Set (CFS) and the Selected Feature Set (SFS) shown in Figure 4. CFS is the original DEAM set, it will remain untouched throughout the whole experiment. SFS is the selected feature set generated by the feature selection algorithm, it will be used to benchmark the model performance in comparison with the CFS.

The CFS will split into training, testing and validation sets during the model evaluation process. They will partition in 70% and 30% ratios. However, the cross-validation will apply, and it can handle how the data will split into the blocks dynamically, which we will mention in detail in the next section.

### 3.5. Feature Selection Process

We mentioned the model complexity issue in feature selection in section, both feature projection and feature selection approaches can use to reduce the feature dimension. However, in comparison to feature projection, the feature selection approach has its advantages and is more like a possible approach to this study. First, it maintained the structure of the original feature after the selection, meaning that we can determine which feature will be useful for the MER, and it is one of the questions we focus on in this paper. Second, the feature selection-based approach makes a more precise task. It provides potential useful features discard from redundant ones on the original feature. This is also the second aim we want to discover in this study hence the Recursive Feature Elimination with Cross-Validation (RFECV) is chosen to address the high feature dimension issue.

RFE is a popular approach in feature selection. It uses a learning model (predictor) recursively removes the weakest features until it reaches the number of specified features, the detail procedure of RFE is show in following:

1. Train the features using multiple predictors
2. Apply the predictor to each feature → obtain its importance
3. For each feature REPEAT until reach to specific number of features has been assessed:
    a. remove the least important feature from the subset
    b. re-evaluate the new feature importance on remain feature subset
4. Finally produces a vector contains all the features with its rank

However, using a plain RFE approach can cause some issues, firstly it requires us to give the algorithm a specific number of features to choose from, but there is no clue about what features are useful for MER in general, hence it's very hard to say which numbers and features are useful. Secondly, the selection of training data makes a high influence on obtaining the ranking of features since the RFE is susceptible to training data. This can cause a high variation when assessing the feature ranking, therefore using the RFE with Cross-Validation (RFECV) can tackle down this problem.

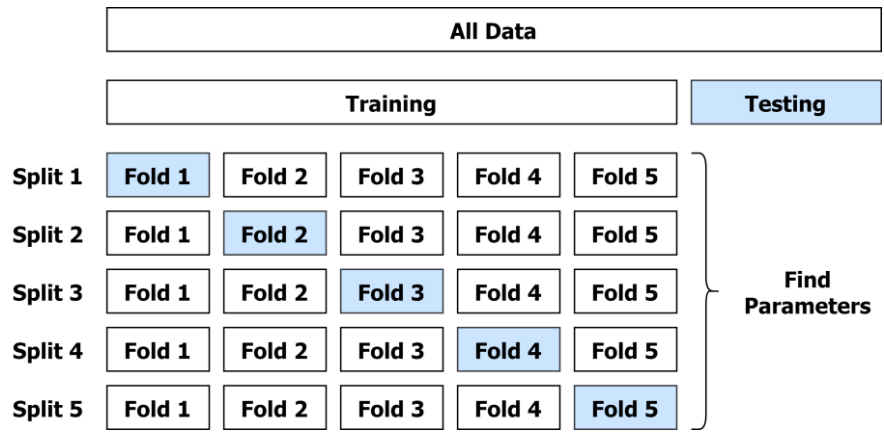

Figure 5. The Cross-Validation Process

Cross-Validation (CV) works by partitioning the training data into the number of smaller subsets evenly. It evaluates the model performances based on how well it deals with the testing data in iterations and keeps track of the model accuracy in all combinations of splitting training and testing data blocks, shown in Figure 5.

Meaning that all combinations of partitioning on the training and testing data will be tested and summarise the result for us at the end, so we do not have to worry too much about determining which data block is the best suited for testing.

Furthermore, there are several advantages over RFE to the RFECV. Firstly, CV can tune the model parameter. It can narrow down finding the optimal number of features by evaluating the blocks in iteration. Secondly, CV can overcome the RFE feature ranking variation by splitting its training data into k different folds (where k is the number of splits). It can produce a more precise evaluation of the feature ranking. Using this technique can enhance RFE and provide a more reliable and robust ranking result.

## 3.6. Model Training Process

We use two regressors to evaluate & benchmark our selected feature set, Support Vector Regression (SVR) and Random Forests (RF). Each regressor was fit with the original full feature and our selected set that we obtained from the previous stage.

SVR has been adapted ubiquitously for the MER regression problem. SVR is robust against overfitting due to its added structural constraints on the discriminant surface. Bai et al. suggested that SVR obtains the best accuracy among the feature extraction methods and has better classification performance for MER.

Fundamentally, SVR finds the optimal linear hyperplane by separating the input data from the feature space. It creates a mapping of the training set into a higher dimensional feature space using a kernel function. Since our approach is based on the regression method, we apply the function which has the most significant deviation from the ground truth $y_s$ for all training data which is as flat as possible in order to get a smooth curve of the function $f(x)$.

$$f(x_s) = m^T \phi(x_s) \tag{1}$$

where m denotes the coefficient of a hyperplane. $x_s$ is the input feature vector, and $\phi$ is the kernel function. We can describe it as an optimization problem defined as below:

$$argmin \frac{1}{2} m^T m + C \sum_{s=1}^{N} (\xi_s + \xi_s^*) \tag{2}$$

The Random Forest is an additive model that makes predictions by combining the decisions of a series from base models. More formally, the definition of it is as below:

$$g(x) = \sum_{i=1}^{n} f_i(x) \tag{3}$$

where the $g(x)$ is the function of the sum of all base models $f_i$. Hence each base model is a decision tree.

Since we are focusing on the dimensional approach, the regression tree is used as the base model. A regression tree predicts the outcome. The model uses the ensemble technique to obtain the predictive performance from the multiple base models. The accuracy is controlled by taking the taking the average value of the predicted value. Since all trees are independent, the regression process of each tree is carried out in parallel. In contrast to the SVR model, the RF model is more efficient in terms of computation in regression problems.

The random forest can calculate the importance of features. It is handy when compared to the SVM model since the feature ranking can be obtained once the training process is finished. There are several metrics available to obtain the feature importance from the model. In our case, we use the Mean Absolute Error (MAE) to do the evaluation, defined as:

$$\frac{1}{N} \sum_{i=1}^{N} |y_i - \mu| \tag{4}$$

where $y_i$ is the valence or arousal annotation from the ground truth, $N$ is the total number of training data, and $\mu$ is the mean of $y_i$. Hence features importance is evaluated by the average reduction of features from all the trees in the random forest.

### 3.7. Model Validation and Benchmark Process

For model validation and benchmark, two commonly used machine learning model, Random Forest (RF) and Support Vector Regression (SVR) is selected, they are benchmarked and tested with our proposed SFS against with CFS under in same hypermeter turning but with the different use of number of features.

The Cross-validation process is also applied in the model evaluation and benchmark stage. CV can provide insightful information about model performance on unseen data and allow us to analyse the model's stability and robustness in a more practical situation against over-fitting. We use ten folds in CV evaluation, meaning that the dataset is partitioned into ten blocks, and model performance is evaluated under ten iterations until every observation of combinations is tested.

The primary goal of CV is to avoid over-fitting, so we don't want to end up with a model which is good at the testing set but bad on unseen data. To do so, CV splits the dataset into the $k^{th}$ folds in equal-sized and iterates testing the model against every possible combination, which is define in follow:

$$CV(\hat{f}) = \frac{1}{N}\sum_{i=1}^{N} L\left(y_i, \hat{f}^{-k(i)}(x_i)\right) \qquad (5)$$

where $CV(\hat{f})$ is the average prediction error sum by each number of $k^{th}$ observations over the total number of $N$ folds in the sample.

## 4. RESULTS

We conducted several experiments to analyze our proposed Selected Feature Set (SFS) from different perspectives. First, we evaluated the raw model performance using our proposed SF. The raw result will be shown in the Model performance result section. Second, we benchmarked our SFS testing against with the baseline CFS by using them in the same hyperparameter turning but different in use of the number of features. The results will be shown in the Model benchmark result section. Lastly, we will collect the selected features proposed by the feature selection algorithm and list them under in the Selected Feature section. We will discuss the details and findings from these results in the next section.

### 4.1. Model Performance Result

The table 2 shows the results of model performance by using our SFS. The SVR and RF models are trained and validated under 10 Cross Validation (CV). The learning models are provided by sklearn framework, the SVR is set in Radial Basis Function (RBF) kernel, 1.0 is set to the regularization and 0.2 is set for epsilon, other parameters are remained by default untouched. The RF model is set in 100 for number of estimators and squared error for criterion. The rest of the parameters also remain default by the framework. The CV is performed in 70% for training, 30% for testing ratio, the score is in the R-Square (R2) metric, and it's calculated using the mean from the 10 CVs results, the detail is shown in Figure 6.

Table 2. Model performance

| Model | Type | Feature Set | Use Features | Score | STD |
|---|---|---|---|---|---|
| Support Vector Regression (SVR) | Valence | CFS | 260 | 0.496 | 0.115 |
| | | SFS | 115 | 0.587 | 0.114 |
| | Arousal | CFS | 260 | 0.502 | 0.122 |
| | | SFS | 74 | 0.645 | 0.117 |
| Radom Forest (RF) | Valence | CFS | 260 | 0.582 | 0.109 |
| | | SFS | 203 | 0.567 | 0.089 |
| | Arousal | CFS | 260 | 0.557 | 0.112 |
| | | SFS | 38 | 0.543 | 0.100 |

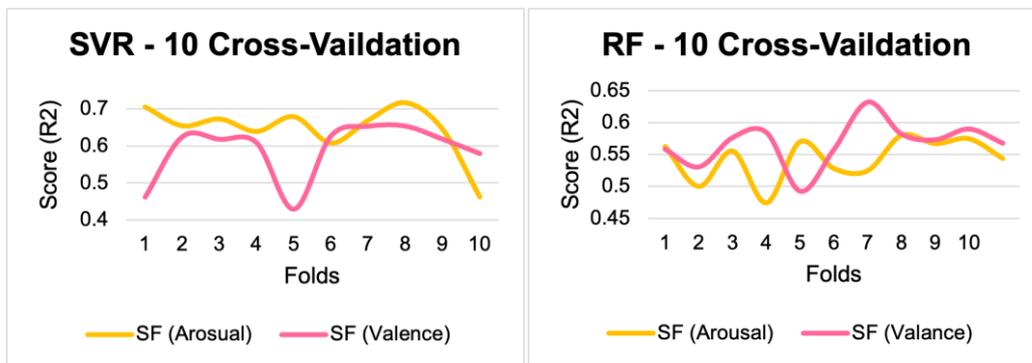

Figure 6. Model Benchmark

## 4.2. Model Benchmark Result

We examined our SFS benchmark against the baseline CFS set, Table 2 shows the detail comparison of SVR and RF models by using these two sets. The SVR and RF model is trained and evaluated by using these two sets under the same hyperparameters. Note that the SVR and RF models have different N Of features for SFS. This is due to the Feature Selection (FS) approach we use in this experiment is wrapper-based; therefore, it relies on a learning model for the proceed the selection process. Both SVR and RF models are considered for the FS learning model, and they produced separate SFS sets based on their models. The Score (R2) is obtained by using the corresponding SFS produced by the FS, and their performance benchmark result is shown in Figure 7. From the benchmark results, our SFS has significantly improved the two model's performance has improved overall.

Additionally, we also summaries the result of Standard Division (STD) for these two models in 10 folds CV process, showing in table G, this result indicates the model stability during the validation in 10 folds. We obtained a tiny boost from STD, SVR about 0.1 in valence, and 0.5% in arousal, RF about 2% in and 1% in arousal.

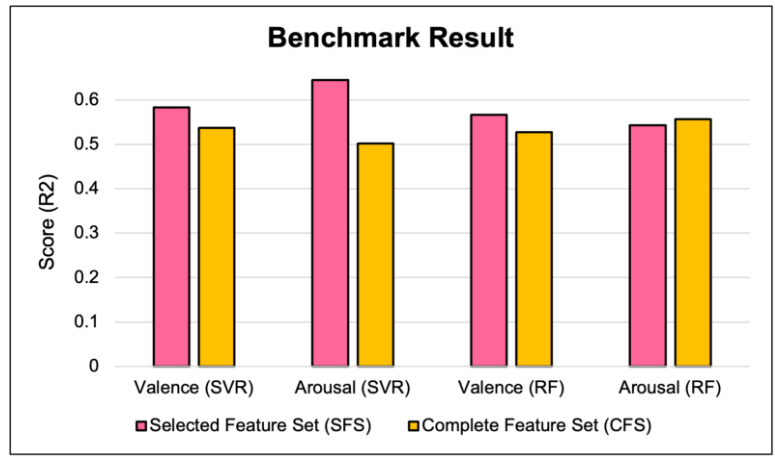

Figure 7.  Benchmark Result Using SFS and CFS

### 4.3. Selected Features Set (SFS)

We want to have a better understanding on relation of features to valence or arousal recognition, hence we collected the data of selected features from the Feature Selection Algorithm (FSA) in order to have a better understanding on them. The SFS for SVR and RF models has been collected separately since the FSA requiring using them separately for the learning model. We performed the several analyses on the features from SFS:

- We analyse the dimension reducing ratio from both models, shown on Table 3, we found the FSA can significantly reduce the features from arousal dimension for both models, SVR reduced from 260 to 74 features (about 71.6%), RF even reduced from 260 to 38 (about 85.4%). The Valence also had a slightly dimension reduce, the SVR model had about 55.8% and RF had about 22%.
- We summaries the number of features selected by these two models by the feature type, we found the top 3 features selected by SVR for valence recognition are audSpec Rfiltsma, pcm fftMag spectral and pcm fftMag mfcc, the top 3 for arousal is audSpec Rfiltsma, pcm fftMag spectral and pcm fftMag fband. For the RF model, the top 3 for valence recognition are audSpec Rfilt sma, pcm fftMag mfcc, pcm fftMag spectral and pcm fftMag mfcc, pcm fftMag spectral and audSpec Rfilt sma for arousal.

Table 3. Feature Dimension Reduce Rate

| Model | Label Axis | N Of Feature Selected | Dimension Reduce |
|---|---|---|---|
| Support Vector Regression (SVR) | Valence | 115 | 55.8% |
| | Arousal | 74 | 71.6% |
| Random Forest (RF) | Valence | 203 | 22.0% |
| | Arousal | 38 | 85.4% |

## 5. DISCUSSION

### 5.1. SVR and RF model performance

In model performance evaluation, we benchmark the performance result of SVR and RF model by using our SFS. we found their results do not make up much different from each other overall. However, The SVR model seems to have done a more outstanding job, outperforming about 10%

Computer Science & Information Technology (CS & IT)                         23better than the RF model in arousal recognition. On the other hand, if we look at Cross-Validation results from both models, the RF model performs a bit more stable than SVR under in 10 CVs, with about 2.5% in valence and 1.7% in arousal. Therefore, selecting on which one of them is a trade-off of performance or stability, and there is no dominant winner or loser in selecting the model for the overall MER task.

**5.2. Selected Features Analysis**

We investigated the selected features from the FSA of both two models. We found no particular dominant features related to the valence or arousal recognition, instead, the best optimal result is obtained by using the features in combinations, however, there are some indication shows that some features combination might be more relevant to certain recognition class, such as both SVR and RF picked up the high amount of spectral, MFCC and spectrum features for the valence, and spectrum, spectral for arousal.

The reason we didn't find any obvious patterns in the relation between the features to arousal or valence might be due to several reasons, first, we treated the emotions as two separate components in valence and arousal rather than classify them into their emotion categories, recognizing a single V or A component can cover the large variety of possible emotion status. Secondly, the nature of the emotion works in a compound way of each other, given a single component like arousal or valence cannot cover the enough information to determine a specific emotion status, such as anger (low valence, high arousal) and sad (low valence, low arousal). Furthermore, even in some extreme case, two diametrically opposites emotion status can be possible in prediction range if we are only specific the valence without arousal e.g., excited (high valence, high arousal) but clam (high valence, low arousal).

**5.3. Selected Feature Set (SFS) Performance**

In order to find out the core motivation of this study, we carry out several analyses about our SFS in the following dimensions:

- We compared the amount of feature reduction selected by Feature Selection Algorithms (FSA).
- We benchmarked the SFS comparing the performance against the CFS using these two models.
- We applied the cross-validation for both models using SFS and CFS to see whether our proposed SFS can benefit the model stability and robustness.

The FSA shows a significant in reducing the feature dimension, SVR reduced from 260 to 74 features (about 71.6%), RF even reduced from 260 to 38 (about 85.4%). The valence also had a slight dimension reduction, the SVR model had about 55.8% and RF had about 22%. Both models got the outstanding number of features reduced, the result indicates the FSA approach is feasible and performs effectively in reducing the features dimension of these two models.

When we put together benchmark of SFS and CFS sets, we can see the model that is using SFS got a significant performance boost, The SVR got an increment about 9.1% in valance and 14.3% in arousal. The RF also got an increase about 3.9% in valence but decrease 1.4% in arousal. Overall, this is a good sign since the SFS uses the less features than the CFS and it can outperform better than use the most available features, this also indicates the idea of "using all existing features may not be efficient for MER tasks due to it creates redundant data lacks informative descriptors related to emotion" is a right way to go [39].

The Cross-validation results of these two models also show similar trends, from the Figure 8 we can see the model that uses SFS performs better overall during the 10 folds, the SVR model shows a significant benefit from SFS, all valence and arousal recognition has boosted. The RF also had similar performances boost overall but sometimes we can see the CFS can perform better sometimes in a particular fold. In order to have a more depth analysis of how the SFS can improve the model stability, we also compute the Standard deviation (STD) of the performance result from 10 CVs. Showing in Table G. We can see we obtained a tiny boost from STD, SVR about 0.1% in valence, and 0.5% in arousal, RF about 2% in valence and 1.2% in arousal. The RF model got more improvement on stability in both valence and arousal recognition tasks.

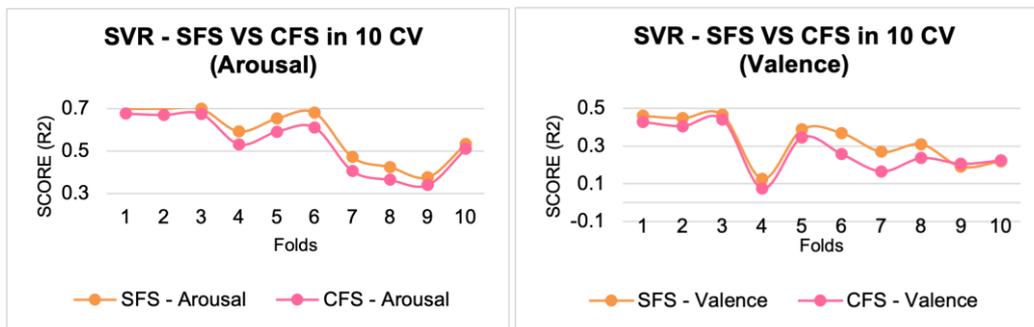

Figure 8. Cross validation result

By putting the CV and the model benchmark results together, we can see there is a sign showing that the irreverent data can cause noise to effect on model stability, using the high amount of audio features doesn't help boost the accuracy of these two models, but also increase the STD during the Cross-validation.

Lastly, we want to discuss some issues about our proposed SFS approach. Firstly, our SFS is not a generic solution because we use the wrapper-based approach for FSA. This type of approach heavily relies on specific a learning model to select features, making the output of the number of features totally different between SVR and RF models. Due to this fact, we can't guarantee our SFS can also perform well in all other machine learning models besides these two, the selected features might have the performance bias issue where it performs outstandingly on one model but is poor on another. Secondly, we also noticed our proposed SFS doesn't obtain a large improvement in the overall model stability. The reason we obtained a tiny bit of boost in STD performance is may due to the loss of data precision since we average out the data from the original set and didn't consider the time variations changes will affect the emotion, but the overall STD results indicate our SFS has the potential benefit to improve the model stability on unseen data. Thirdly we treat the valence and arousal separately as the individual regressor for the feature selection and recognition task, there might be a compound relation between features to valence and arousal, in this case, a hybrid feature selection approach may be the key to solving this problem.

## 6. CONCLUSION AND FUTURE WORK

In this work, we investigated whatever the feature selection algorithm (FSA) can have the benefit to MER performance. First, we performed the FSA on the Complete Feature Set (CFS) to propose our Selected Feature Set (SFS). Second evaluated our SFS performance by training it on two popular machine learning models Random Forest (RF) and Support Vector Regression (SVR) and benchmark them again to the CFS. Third we applied the Cross-Validation (CV) analysis to these models to find out how our SFS can impact the model stability in both arousal and valence reorganization tasks. Finally, we collected the selected features from the FSA and attempted to find out the relationship between the audio features to an emotion status.

Our results SFS show the use of FSA is a right direction to go to improve the overall MER performance, the FSA can not only significantly reduce the feature dimension, improve the model performance but also has the potential to enhance the stability on unseen data. Furthermore, we found there is no dominant feature particularly useful for the MER, the best result is by using them in combination.

However, we also noticed there are some limitations is in our proposed approach. Firstly, our SFS is not generic enough, the use of wrapper-based FSA is heavily relying on a learning model to select features, this may cause bias on the certain machine learning model, in future, a more generic FSA approach may need to be consider, such as filter-based FSAs, hybrid FSAs. Secondly, the use of data conversations from time-sequence to average mean can cause information loss, a time series forecasting model can consider using in further research, e.g., seasonal support vector regression (SSVR). Lastly splitting valence and arousal into two different regressions may not fully cover the relation from the features to an emotion status, since the nature of emotion may have a compound relationship to each other, therefore the hybrid-based FSA approach may consider for the future research.

## 7. DATA AVAILABILITY

The dataset described in this article is openly available in repository hosted by University of Geneva: https://cvml.unige.ch/databases/DEAM

## AUTHORS

**Name:** Le Cai

Le Cai is a Ph.D. student at Faculty of Engineering and Information Technology, University of Technology Sydney, His research interests include Emotion Recognition, Music information Retrieval and Data mining.

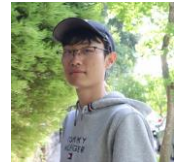

**Name:** Sam Ferguson

Sam Ferguson is a Senior Lecturer within the School of Computer Science at UTS. He focuses on sound and music and its relationship with creativity and human experience. He has more than 80 publications in areas as diverse as spatial hearing and loudness research, to data sonification, emotion, and tabletop computing.

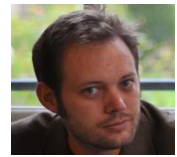

**Name:** Hai Yan (Helen) Lu

Dr Haiyan (Helen) Lu is an Associate Professor in the School of Computer Science, Faculty of Engineering and Information Technology, University of Technology Sydney (UTS). She is a core member of the Decision Systems and e-Service Intelligence Lab in the Centre for Artificial Intelligence (a part of former Centre for Quantum Computation and Intelligent Systems) at UTS.

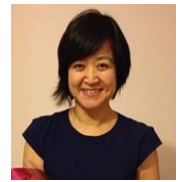

**Name:** Gengfa Fang

Gengfa Fang received a Ph.D. degree in wireless communications from the Institute of Computing Technology, Chinese Academy of Sciences, Beijing, China, in 2007. From 2009 to 2015, he was with the Department of Engineering at Macquarie University, Sydney. In 2016, he moved to the University of Technology Sydney, Ultimo, NSW, Australia, where he is currently a Senior Lecturer.

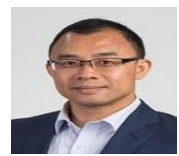